\documentclass[12pt,epsfig]{article}
\usepackage{hyperref}
\usepackage{graphicx}
\usepackage{amssymb}

\newcommand{\beq}[1] {\begin{equation}\label{#1} }
\newcommand{\eeq} {\end{equation} }

\newcommand{\bea}[1]{\begin{eqnarray}\label{#1} }
\newcommand{\eea}{\end{eqnarray}}

\newcommand{\vn}{{\vec{n}}}

\newcommand{\si}{\sigma}
\newcommand{\hmu}{{\hat\mu}}
\newcommand{\hnu}{{\hat\nu}}

\newcommand{\hh}{{\hat{h}}}

\newcommand{\tih}{{\widetilde{h}}}
\newcommand{\tp}{{\widetilde\phi}}
\newcommand{\tA}{{\widetilde{A}}}

\newcommand{\der}{\partial}

\begin{document}

\vspace*{-0.5cm}
\begin{flushright}
OSU-HEP-05-12\\
SU4252-817
\end{flushright}
\vspace{0.5cm}

\begin{center}
{\Large {\bf Monojet and Single Photon Signals \\
from Universal Extra Dimensions} }

\vspace*{1.5cm}
 C. Macesanu\footnote{Email address: cmacesan@physics.syr.edu}$^{,\dag}$,
S. Nandi\footnote{Email address: s.nandi@okstate.edu}$^{,\ddag}$
and C. M. Rujoiu\footnote{Email address:
marius.rujoiu@okstate.edu}$^{,\ddag,*}$

\vspace*{0.5cm}

$^{\dag}${\it Department of Physics, Syracuse University\\
Syracuse, New York 13244}

$^{\ddag}${\it Department of Physics, Oklahoma State University\\
Stillwater, Oklahoma, 74078\\}

$^{*}${\it Institute of Space Sciences, Bucharest-Magurele\\
Romania, 76900\\}
\end{center}

\begin{abstract}
The usual universal extra dimensions scenario does not allow for
single production of first level Kaluza-Klein (KK) excitations of
matter due to the KK number conservation. However, if the matter
fields are localized on a fat brane embedded in a higher
dimensional space, matter-gravitation interactions violate KK
number, and the production of single KK excitations becomes
possible. In this paper we analyze the production of a single KK
matter excitation together with a graviton in the final state, and
study the potential for discovery at the Tevatron and Large Hadron
Collider.
\end{abstract}

\section{Introduction}

Some of the more interesting developments in modern particle
physics are based  on
the idea that our universe
has more dimensions than the four already known.
Moreover, there may be reasons to believe that the size of these
extra dimensions is rather large \cite{anton,add}
(maybe as large as inverse eV size),
since then one could understand the weakness of gravitational
interaction in contrast with the other fundamental forces.  This in turn can
have interesting implications for the phenomenology of present day and
near-future colliders.

 The experimental signatures of such models depend on the
fields which  propagate in extra dimensions (the bulk).
In the simplest case, the ADD scenario \cite{add},
only gravity propagates in the extra dimensions.
 Our universe is viewed  as a 4D-brane embedded
in a bigger 4+$N$ dimensional space. In this picture, matter
fields and gauge bosons are confined to the brane, gravity is
naturally weak because it is propagating in $N$ extra compact
dimensions, and the hierarchy problem is resolved by bringing the
fundamental scale of gravity close to the electroweak scale
according to the formula:
\beq{grv-rel} M_{Pl}^2 = M_D^N \left({r \over 2 \pi}\right)^{N+2}.
\eeq
Here $M_{Pl} \simeq 10^{19}$ GeV is the  Plank scale in 4 dimensions,
while  $M_{D}$
 is the new fundamental scale of gravity in 4+$N$ dimensions. One can
see that with $M_{D}$ of order TeV, one would obtain $r$ as large as
 eV$^{-1}$ (for $N=2$) up to MeV$^{-1}$ (for $N=6$). A
general feature of this type of models is the existence of the KK
towers of excited states of the graviton. The mass splitting
between levels is proportional to the inverse of the compactification
radius $\Delta m = 2 \pi/r$.

One can construct extensions of the above  model where matter
fields also propagate in extra dimensions \cite{others}. For the
case when all matter fields propagate in the bulk, one obtains the
universal extra dimensions scenario (UED) \cite{acd}. This
scenario has several interesting features, which derive from the
existence of a selection rule that requires KK number conservation
for interactions. This means that the KK excitations have to be
pair-produced, and, moreover, they cannot decay directly to the SM
particles. The resulting phenomenology is then somewhat similar to
that of supersymmetric theories \cite{CMSexp} with almost
degenerate superpartner masses (the mass splittings between the
first lever KK excitations being generated by radiative
corrections \cite{CMS}). The lightest KK particle (LKP) is similar
to the lightest supersymmetric particle (LSP) and is a possible
dark matter candidate \cite{ST}.

Although interesting, the simplest UED scenario described above
leaves some questions unanswered. Since experiments have not
detected any KK excitations of matter yet, the radius of the extra
dimensions in which matter propagate must be at least of order of
TeV$^{-1}$.
 But this would spoil the features the gravity sector
had in the ADD-like models and would reintroduce a hierarchy
between the electroweak and the gravity scale. Of course, one can
introduce asymmetric models containing sub-millimeter size extra
dimensions in which only gravity propagates and TeV$^{-1}$ size
extra dimensions in which only matter propagates (see, for
example, \cite{asym}). But we will not pursue that possibility
here.

In this work, we consider the case in which the gravity sector has
the same features as the ADD models (sub-millimeter size compact
extra dimensions), but matter can propagate only distances of
order TeV$^{-1}$ along them. This is the fat-brane scenario
\cite{rigolin}, which preserves the ADD solution to the hierarchy
problem, as well as give rise to new
interesting low-energy phenomenology associated with production of
KK matter excitations. Having UED on a fat brane allows
gravity-matter interactions to break the KK number conservation.

As a consequence, the KK excitations of matter can decay via
gravitational radiation (massless graviton as well as KK graviton
emissions). If the LKP is the excitation of the photon, this will
lead to signals with two high transverse momentum photons and
missing energy in the final state \cite{cmn2}. Also, although the
gravitational decay width of these particles is very small when
considering only one graviton, the total width increases
substantially when one considers the contribution from the large
number of gravitons with masses from an eV (or MeV) al the way to
a TeV. In fact, the gravitational decay widths may be large enough
so that the KK excitations decay mostly to SM quarks and gluons
plus gravitons, rather than to the LKP. The experimental signature
in such a case, assuming KK pair-production, would be two jets
plus missing energy. The resulting phenomenology has been studied
in \cite{mmn}.

A second consequence is  that it is also possible to produce a
single KK excitation of matter (unlike the usual UED case, where
these excitations are produced in pairs). This can take place in
two ways; either through the exchange of virtual gravitons, or
with the production of a real graviton in the final state. In the
first case, a SM quark or gluon is produced together with a KK
excitation in the final state; the expected signals for such
processes are two large $p_T$ jet events plus missing energy
\cite{grv-cos}. In the second case, which constitute the topic of
the present paper, one will have a graviton in the final states
together with a single KK excitation or a SM quark or gluon. This
type of processes has as signature monojet events (or single
photon, if the KK excitation decays to the LKP first),
 with large missing energy.

The outline of the paper is as follows. In the next section we
will give a brief overview of the model used, comprising the
matter and gravity sectors. In section 3 we comment on some
interesting features of the  production cross-sections for the
 relevant processes. One such feature is an enhancement of the
cross section for very light gravitons in the final state. In
section 4 we discuss the phenomenological signals for the
production of one graviton plus one excited KK state (or a SM
quark or gluon) in the UED model with a fat brane, at the Tevatron
and LHC. We end with conclusions.

\section{Model description}

In our scenario, matter propagates on a fat brane with four
Minkowsky plus one compact extra dimensions. The length scale $R$
of the compact dimension is of order TeV$^{-1}$. In order to have
as zero modes only the SM content and to project out all other
additional (unwanted) zero modes, we impose a $S_1/Z_2$ orbifold
symmetry. This fat brane lives in a higher dimensional space (the
bulk) in which only gravity propagates; the radius of these extra
dimensions can be as large as eV$^{-1}$, and it is related to the
4D Planck mass $M_{Pl}$ by the ADD relation (\ref{grv-rel}).

The $4+N$ dimensional graviton is expanded in KK modes\cite{hlz}:
\beq{grav_exp} \hh_{\hmu\hnu}(x,y)\ =\ \sum_{\vec
n}\hh_{\hmu\hnu}^{\vec{n}}(x)\\exp\left(i {2\pi
\vec{n}\cdot\vec{y}\over r}\right)\ .
\eeq
The `hat' denotes quantities which live in 4+$N$ dimensions:
$\hmu, \hnu = 0,\ldots,3,5,\ldots 4+N$, while $\mu, \nu =
0,\ldots,3$. At each KK level we have the decomposition of the
$\hh_{\hmu\hnu}$ field  into 4D tensor
$h_{\mu\nu}$, $N$  vectors $A_{\mu i} $  and
$N(N+1)/{2}$ scalar fields $\phi_{ij}$ by:
\beq{hh_def} \hh_{\hmu\hnu}^{\vec{n}}\ =\
V_N^{-1/2}\left(\begin{array}{cc}
h_{\mu\nu}^{\vec{n}}+\eta_{\mu\nu}\phi^{\vec{n}} & A_{\mu i}^{\vec{n}}\\
A_{\nu j}^{\vec{n}}   &  2 \phi_{ij}^{\vec{n}}
\end{array}\right)\
\eeq
where $V_N= r^N$ is the volume of the $N$-dimensional torus. Not all these
fields are independent; by imposing the de Donder gauge fixing condition
\bea{constr}
\partial^{\hmu}(\hh_{\hmu\hnu}-\frac{1}{2}{{\hat\eta}}_{\hmu\hnu} \hh)=0 \
 ,\eea
together with $n_{i}A_{\mu i}^{\vec{n}}=0,\ n_{i}\phi_{ij}^{\vec{n}}=0$,
one can eliminate the spurious degrees of freedom and express the Lagrangian
at each KK level $\vn$
in terms of one physical massive spin 2 field $\tih_{\mu \nu}^\vn$,
$N-1$ massive vector gravitons $\tA_{\mu i}^\vn$ and $N(N-1)/2$ massive scalars
$\tp_{ij}^\vn$. The vector and scalar physical fields also satisfy
$n_i \tA_{\mu i}^\vn = 0,\  n_i \tp_{ij}^\vn = 0 $.

Where the gravity is not affected in any way by orbifolding, the
matter fields are.
The corresponding decompositions for fermions,
scalars and gauge bosons are:
\bea{sm_fer} Q  &  = & \frac{1}{\sqrt{\pi R}} \left\{ Q_{L} +
\sqrt{2} \sum_{n=1}^{\infty} \left[ Q_L^n  \cos \left(\frac{n
y}{R} \right)
 + Q_R^n  \sin \left(\frac{n y}{R} \right) \right] \right\}, \nonumber \\
q  &  = & \frac{1}{\sqrt{\pi R}} \left\{ q_{R} + \sqrt{2}
\sum_{n=1}^{\infty} \left[ q_R^n  \cos \left(\frac{n y}{R} \right)
 + q_L^n  \sin \left(\frac{n y}{R} \right) \right] \right\},\nonumber\\
(\Phi, B_{\mu}^a)  & = & \frac{1}{\sqrt{\pi R}}\left[ (\Phi_0,
B_{\mu, 0}^a)  + \sqrt{2} \sum_{n=1}^{\infty} (\Phi_n, B_{\mu
,n}^{a})  \cos(\frac{n y}{R}) \right]. \eea
Here $Q(q)$ are the 5D fermionic doublets (singlets) under SU(2) whose
zero mode are the usual SM fermionic doublets (singlets); $\Phi$ is the
scalar field (Higgs) and $B^{a}_{\mu}$ are the vector (gauge) fields.
We work in a gauge where $B^{a}_5$ = 0 \cite{Dienes, Santamaria}.
The fields in
(\ref{sm_fer}) have the following parities under $Z_{2} (y
\rightarrow -y)$:
\beq{} Q_{L}(x,y)=Q_{L}(x,-y),\ \ Q_{R}(x,y)= -Q_{R}(x,-y),\ \
B^{a}_{\mu}(x,y)=B^{a}_{\mu}(x,-y). \eeq
The effective 4D Lagrangian and the Feynman rules for the interactions of KK
excitations has been discussed in \cite{mmn}.
The tree level masses of the particles in the SM fields' towers are multiples
of $M = 1/R$. The interactions of the KK excitations are similar to those of
the SM partners; with the obvious caveat that in the case of electroweak
interactions, the fermion fields in the left
handed doublet $Q_L$ couple to the electroweak gauge fields as pure $SU(2)_L$
doublets, while the $Q_R$ fermion fields couple only to the hypercharge
$U(1)_Y$ gauge field. Moreover, the interactions between matter fields obey
KK number conservation rules (at tree level), which requires
that KK particles be produced in pairs at colliers
(see for example \cite{acd,mmn}).

The interactions between matter and gravity are obtained from the
following $4+N$ dimensional action:
\beq{act_gen} {\cal S}_{int} = -{{\hat \kappa} \over 2} \int d^{4+N} x
\ \delta(x^6) \ldots \delta(x^N)\
 \hh^{\hmu\hnu} T_{\hmu\hnu}
\eeq
with $T_{\hmu\hnu}$ the energy-momentum tensor of the 5D matter
and ${\hat \kappa}$ the strength of the $4+N$ gravitational
coupling (related to the 4D one by $\kappa = {\hat \kappa}
V_N^{-1/2}$). To obtain the effective 4D Lagrangian, one needs to
expand the fields in (\ref{act_gen}) in KK modes and integrate
over the extra dimensions (which essentially means the fifth
dimension, due to the delta functions). More details can be found
in \cite{grv-pap}. The Lagrangian and resulting Feynman rules for
matter-gravitation interactions are similar to those obtained in
\cite{hlz} for the case of matter propagating into four
dimensions; one has extra couplings to the vector and scalar
gravitons due to components of the matter energy-momentum tensor
involving the fifth dimension: \bea{l_4d2} {\cal L}_{int} & = &
-{\kappa \over 2} \sum_{\vn} \left\{ \left[ \tih^\vn_{ \mu \nu} +
\omega \left( \eta_{\mu \nu} + \frac{\der_\mu \der_\nu}{m_\vn^2}
\right) \tp^\vn \right]
T_{n_5}^{\mu \nu}   - \right.  \nonumber \\
& & \left. 2 \tA^\vn_{\mu 5}
T_{n_5 5}^{~\mu}  +
 \left(  \sqrt{2} \tp^\vn_{55} - \xi \tp^\vn \right) T^{n_5}_{55}
\right\} \ , \eea
with $\omega = \sqrt{2/[3(N+2)]}$.
Moreover, the interaction vertices are  multiplied by form-factors
\beq{form_fact}{\cal F}^{(c,s)}_{lm|n} \sim { 1\over \pi R}
\int_0^{\pi R} dy (\cos,\sin) \left({l y \over R}\right)
 (\cos,\sin) \left({m y \over R}\right) \exp \left(2 \pi i {n y \over r}\right)
\eeq
(here $l$ and $m$ are the KK numbers of the matter excitations, while $n =n_5$
is the fifth component of the graviton KK number $\vn$),
which describe the overlap of the graviton and matter wave functions
in the fifth dimension. As we shall see in the following, these form-factors
play an important role in the production of KK excitations of matter.

The phenomenology of the UED model on the fat brane is
significantly affected by the inclusion of the gravitational
interactions. In absence of gravity, KK number conservation at
tree level \cite{acd} implies that the first level KK excitations
are stable (since they are also degenerate in mass). Radiative
corrections \cite{CMS} will break the degeneracy, and introduce
boundary terms violating KK number conservation; however, KK
parity is still conserved, and the lightest KK particle (LKP) is
stable. The phenomenology of such a model \cite{CMSexp} results
from the pair production of KK excitations of quarks and gluons in
hadron colliders, which will then decay to LKP (which is the
photon excitation $\gamma^*$) radiating neutrinos, leptons and/or
quarks. The LKP, being weakly interacting and stable, will appear
as missing energy in the detector. The observable jets or leptons
have an upper limit on their energy given by the split between the
KK masses of the initial quark or gluon excitations and the
$\gamma^*$, therefore, they can be hard to see. The resulting
signal is at first glance similar to that of a supersymmetric
model with almost degenerate superpartner masses, where the
lightest supersymmetric particle is a neutralino.

Introducing gravitational interactions affects the phenomenology
in the following way. First it can modify the decay pattern of KK
excitations: KK quarks and gluons produced at a hadron collider
can decay directly to SM quarks and gluons plus gravitons. The signal in
this case will be two jets plus missing energy (associated with
the graviton) \cite{mmn}. Depending on the relative magnitudes of
the gravitational decay widths compared to the decay widths
between first level KK excitations of matter, it is also possible
that the KK quarks and gluons will first decay to the LKP, which
then will decay gravitationally; the signal in this case will be
two high $p_T$ photons plus missing energy (and several soft jets
and/or leptons from the primary decays of the $q^*, g^*$s)
\cite{cmn2}. Second, one can now have production of a single KK
excitation of matter, mediated by gravity. The case when the
gravitons involved in production are virtual particles has been
discussed in \cite{grv-cos}; in this article we will discuss the
case when gravitons appear as final state particles.


\section{Processes with final state gravitons}

\begin{figure}[t!] 
\centerline{
   \includegraphics[height=1.in]{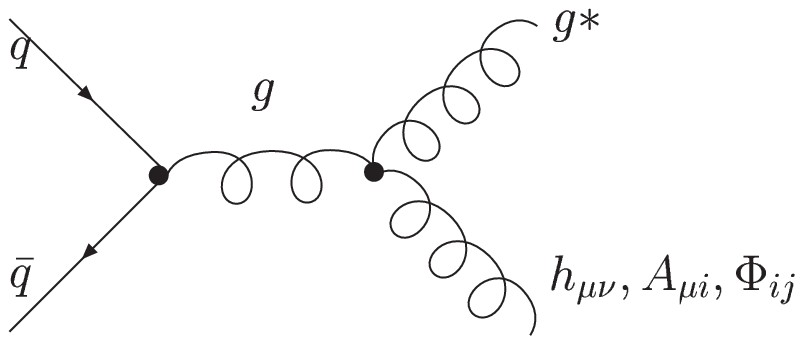}\hspace{2.cm}
   \includegraphics[height=1.in]{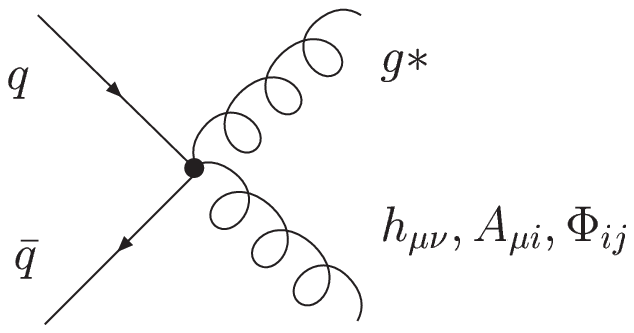}
}\vspace{1.cm}
\centerline{
   \includegraphics[height=1.5in]{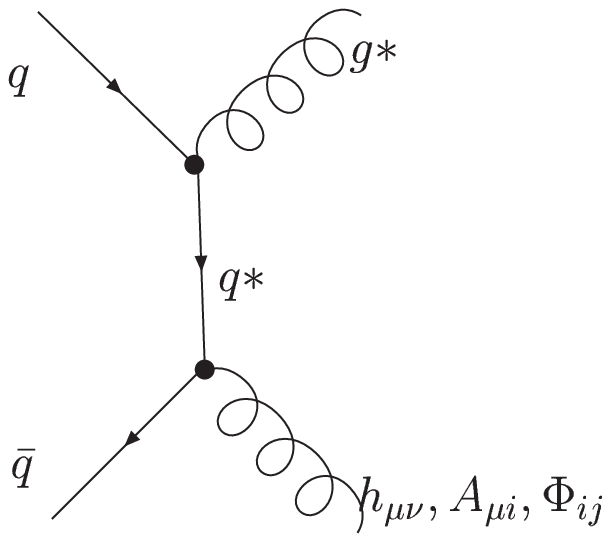}\hspace{2.8cm}
   \includegraphics[height=1.5in]{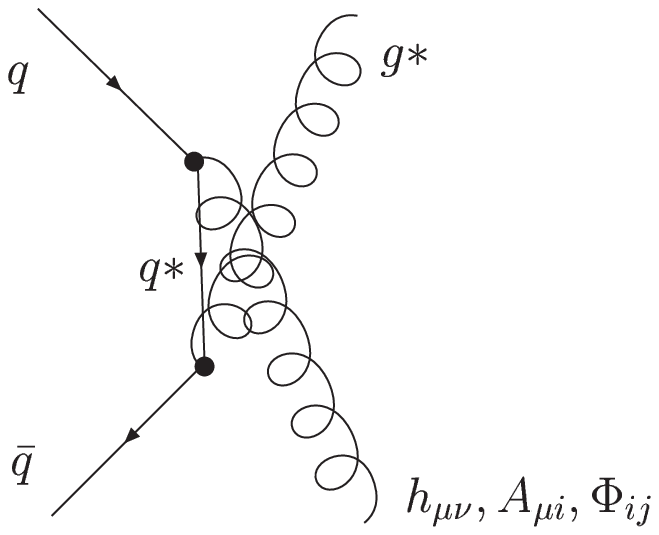}
}
\caption {Feynman diagrams contributing to the production of a
$g^*$ and a graviton KK excitation  (either $h_{\mu\nu},A_{\mu i}$ or
$\Phi_{ij}$)
at hadron colliders.}
\label{fey_diag}
\end{figure}

In this section we will discuss some characteristic features of
hadron collider production of KK gravitons and either
Standard Model particles or their KK excitations. A typical set of Feynman
diagrams contributing to these processes is shown in Fig. \ref{fey_diag};
these correspond to the case of a final state gluon excitation and a KK
graviton.
Similar diagrams contribute for the SM gluon production, but only with the
spin 2 gravitons $h^{\mu \nu}$ in the final state (since the SM particles do
not couple to the vector gravitons, and their coupling to the scalar
gravitons is proportional to their mass, and therefore quite small).

Let us first discuss the production of SM matter particles in the
final state. The amplitudes then are similar to those evaluated in
the pure ADD case \cite{GRW,peskin-add} (when matter lives on the
4D brane), with the only difference being the appearance of the
thick brane form-factor \beq{SM_prod} \sigma = |{\cal
F}^{c}_{00|n_5}|^2 \ \sigma_{ADD} \ , \eeq with \cite{grv-pap}:
\beq{ff_00} {\cal F}^c_{00|n_5} = { 1\over \pi R} \int_0^{\pi R}
dy \ \exp \left( 2 \pi i n_5 y \over r\right) \ = \ i  {e^{i x } -
1  \over x} \ .\eeq In the above expression, we have defined $x$
as $x = \pi m_5 R $, where $m_5 = 2 \pi n_5/r$ is the graviton
momentum along the fifth dimension. Note that the absolute value
of the form-factor is smaller than one, so its contribution has
the effect of multiplying the total cross-section (obtained after
adding the contributions of all the gravitons in the KK tower) by
a parameter $r < 1$. However, this is not a big effect. By
numerical simulations, we find that the parameter $r$ takes values
roughly in the interval $1$ (corresponding to the case $1/R \gg
M_S$, where $M_S$ is the upper limit on the graviton mass
contributing to the process\footnote{While theoretically all
gravitons in the KK tower can contribute to the process, there is
an upper limit on the graviton mass imposed by the collider energy
and luminosity considerations.}) to around 0.7 (corresponding to
the case when $1/R $ is of the same order of magnitude as $M_S$).

At first glance this might appear counterintuitive. Indeed, we
know that for more that two extra dimensions, generally heavy
gravitons will bring the dominant contribution to the total
cross-section \beq{tot_cs} \sigma^T \ = \ \sum_\vn \sigma^\vn \ =
\ \frac{M_{Pl}^2}{M_D^{N+2}} \int m_h^{N-1} d m_h \ d \Omega_N \
\sigma^\vn \eeq (where we have replaced the sum over graviton
states by an integral \cite{hlz}), due principally to the fact
that the density of graviton states increases as $m_h^{N-1}$. At
large masses, the form-factor (\ref{ff_00}) behaves like $1/m_5$,
therefore one should expect to get substantial reduction in the
total cross-section. However, while it is true that most of the
cross-section comes from large values for $m_h = 2\pi
\sqrt{\vn^2}/r$, that is for $m_h$ of order $M_s$, this actually
corresponds to significantly smaller values\footnote{For example,
if we assume a flat distribution for the cross-section over a
sphere of radius $M_S$ in $N$ dimensions, the average value for
$m_5$ would be of order $M_S/N^2$.} of $m_5 = 2\pi n_5 /r$. As a
consequence, most of the gravitons contributing to the total cross
section have $ x =\pi m_5 R \ll 1$, and therefore $|{\cal
F}^{c}_{00|n_5}|^2 \sim 1$.

Let us consider next the production of gravitons together with the
KK excitation of a quark or gluon in the final  state. Let us for
now assume that the $q^*$ or $g^*$ will decay to a Standard Model
quark or gluon by radiating a graviton. Then, the phenomenological
signal will be a jet plus missing energy (carried away by the two
gravitons in the final state). So at first order the signal is
indistinguishable from that coming from the production of a SM
quark or gluon and a graviton. Moreover, since in this case we
have to produce also a massive particle in the final state (the KK
excitation of matter) one might think that this type of process
will not give a significant contribution.

However, this conclusion is hasty. To see that, let us consider again the
cross-section for the production of a SM particle with a graviton of mass
$m_h$. From dimensional analysis (see also \cite{peskin-add}),
 one can estimate this cross-section to be of order
\beq{sm_cr_es} \sigma_{SM} \ \sim \ \frac{\alpha_s}{M_{Pl}^2}
\left( 1 + {\cal O}(\frac{m_h^2}{s})\right)\ .\eeq
(assuming here $m_h^2 \ll s$)
Let us consider the case of a KK matter particle of mass $M$
in the final state. A naive
estimate will give an expression of the form (\ref{sm_cr_es}), with terms
of order $M^2/s$ in the final state. However, if one evaluate the amplitude
squared for the process with the spin 2 graviton in the
final state, one obtains
\beq{kk_cr_es} \sigma_{KK} \ \sim \ \frac{\alpha_s}{M_{Pl}^2}
\left(\frac{M}{m_h}\right)^4
\left( 1 + {\cal O}(\frac{m_h^2}{s}) + {\cal O}(\frac{M^2}{s}) + \ldots
\right)\ .\eeq

The term $(M/m_h)^4$ can lead to a great enhancement of the
cross-section for producing light gravitons in the final state. To
see how big this enhancement is, one just need remember that $m_h$
can be as low as eV, while $M$ is of order TeV; this leads to a
$10^{48}$ factor. The appearance of this enhancement factor is due
to the breaking of translation invariance in the 5th dimension by
the brane. This has as consequence the non conservation of 4D
energy-momentum tensor of the matter $k^\mu T_{\mu \nu} \neq 0$,
for interactions involving matter excitations with different KK
numbers. Instead, one has $k^M T_{MN} =0$ (with the indices $M,N$
going from 0 to 5), or
  $$ k^\mu T_{\mu \nu}  \ = \ -k^5 T_{5\nu} \ \sim \ \Delta m_{kk} \ ,$$
where we used the fact that the momentum in the fifth dimension is proportional
to the KK mass. In our case, we have one SM particle becoming a first level
excitation, therefore $\Delta m_{KK} = 1/R = M$. Then, if one considers
the amplitude for the creation of a KK matter excitation by the radiation
of a graviton with momentum $k$:
$$
\sum_{\hbox{spin}}|{\cal M}(q \rightarrow q^* h^\vn)|^2 \ \sim \
\langle q^* | T_{\mu \nu} | q \rangle \langle q^* | T_{\rho \si} | q \rangle\
B^{\mu \nu , \rho \si}(k) \ ,
$$
one sees that due to terms $\sim k^\mu k^\nu k^\rho k^\si /m_h^4$
in the graviton polarization sum $B^{\mu \nu , \rho \si}(k)$ (see,
for example, \cite{hlz} for the full expression), one will obtain
$ \sum_{\hbox{spin}}|{\cal M}(q \rightarrow q^* h^\vn)|^2 \ \sim \
({M}/{m_h})^4 $. Similar behavior holds for the production of a
scalar graviton in the final state (although in this case terms
$\sim k^\mu k^\nu/m_h^2$ are due to the $\der^\mu
\der^\nu/m_{\vn}^2$ factor in the interaction lagrangian
(\ref{l_4d2}) rather than to the sum over polarizations),
 while for the case with a vector graviton
field $\tA^\mu$ in the final state, one has
$\sigma \sim \alpha_s/M_{Pl}^2  \ (M/m_h)^2$.

A behavior of the production cross-section $\sim ({M}/{m_h})^4$
would mean that for the case of $N=2,3$ one would be able to probe
very large values of $M_D$ (by contrast, the total production
cross section for large $N$ values is not affected very much,
since, as we have mentioned above, the contributions of heavy
gravitons are enhanced by a density of states factor $m_h^{N-1}$,
which will win over $(1/m_h)^4$ factor). However, we still have to
take into account the form factors describing the overlap of
graviton and matter wave functions on the brane. For the processes
with the spin-2 graviton or the scalar gravitons in the final
state, the form factor multiplying the production cross-section
will be $|{\cal F}^c_{01|n_5}|^2$, while for the vector graviton
is $|{\cal F}^s_{01|n_5}|^2$. Here \beq{prod_ff} {\cal
F}^{(c,s)}_{01|n_5} \ = \ \frac{\sqrt{2}}{\pi R} \int_0^{\pi R}
 dy \  (\cos, \sin) \biggl( { y\over R}\biggr) \
 \exp\biggl( 2 \pi i{n_5 y \over r}\biggr) \ ,
\eeq
and therefore
\beq{dec_ff1}
|{\cal F}^c_{01|n_5} |^2\ = \
\frac{4 x^2}{(\pi^2-x^2)^2}\left[1+\cos(
x)\right] \ , \ \
 |{\cal F}^s_{01|n_5} |^2\ = \ \pi^2 \frac{|{\cal F}^c_{01|n_5} |^2}{x^2} \ ,
\eeq
with $x$ as defined after Eq. (\ref{ff_00}). According to the discussion above,
generally small values of $x$ are relevant for the total cross-section; we then
have
\beq{small_xff}
|{\cal F}^c_{01|n_5} |^2 \ \sim \ \frac{8 x^2}{\pi^4} \ = \
 \frac{8}{\pi^2} \frac{m_{5}^2}{M^2}
 \ , \ \
|{\cal F}^s_{01|n_5} |^2\ \sim \ \frac{8}{\pi^2} \ ,
\ \ \hbox{for} \ x \ll 1 \ .
\eeq

We see then that the form-factors contribute additional terms of
order $(m_h/M)^2$ to the cross-section for the production of
spin-2 and scalar gravitons. For a small number of extra
dimensions ($N=2,3$) this has an effect of making the enhancement
factor in front of the cross-section the same for all final
states, roughly $(M/m_h)^2$. This will enhance the cross-section
somewhat, but not very much. (If one takes into account the
density of graviton states $m_h d m_h$ for $N=2$ , one sees that
this is a roughly logarithmic effect).  In fact, we find that the
cross-section for production of gravitons with a KK excitation is
still smaller than the production of gravitons with SM particles
(although typically lighter gravitons are predominant in the first
case). For $N=4$ to 6, the form factor has a net effect of
reducing the contributions coming from spin 2  and scalar
gravitons in the final state (unlike the enhancement due to
breaking of 5D translational invariance, which is important mostly
for light gravitons, the form-factor cancellation effect is valid
for large graviton masses as well). As a consequence, we find that
for such processes, the gravitons appearing in the final state for
values of $N$ larger than 4 are mostly vector gravitons. By
contrast, for $N=2,3$ generally final states with spin-2 gravitons
will dominate.

\section{Results}

In this section we present results for phenomenological signals at
Tevatron Run II and LHC due to the production of an SM particle and/or a
KK excitation together with a
KK graviton in the final state.

We start by considering the monojet plus missing energy signal.
 There are two contributions to this signal: first comes from the
production of a SM quark or gluon in the final state. The second contribution
comes from the production of a KK excitation of a quark and gluon, which
subsequently will decay by radiating a graviton to a SM particle.
As mentioned in the previous
section, the first type of contribution is generally dominant.
The signal is then
roughly the same one would obtain in a pure ADD theory (with matter stuck on
a 4D brane). There is a small difference for our model due to the appearance
of the form factor associated with matter propagating in the 5th dimension;
however, the numerical importance of the form factor is small.

\begin{figure}[t!] 
\centerline{
   \includegraphics[height=3.in]{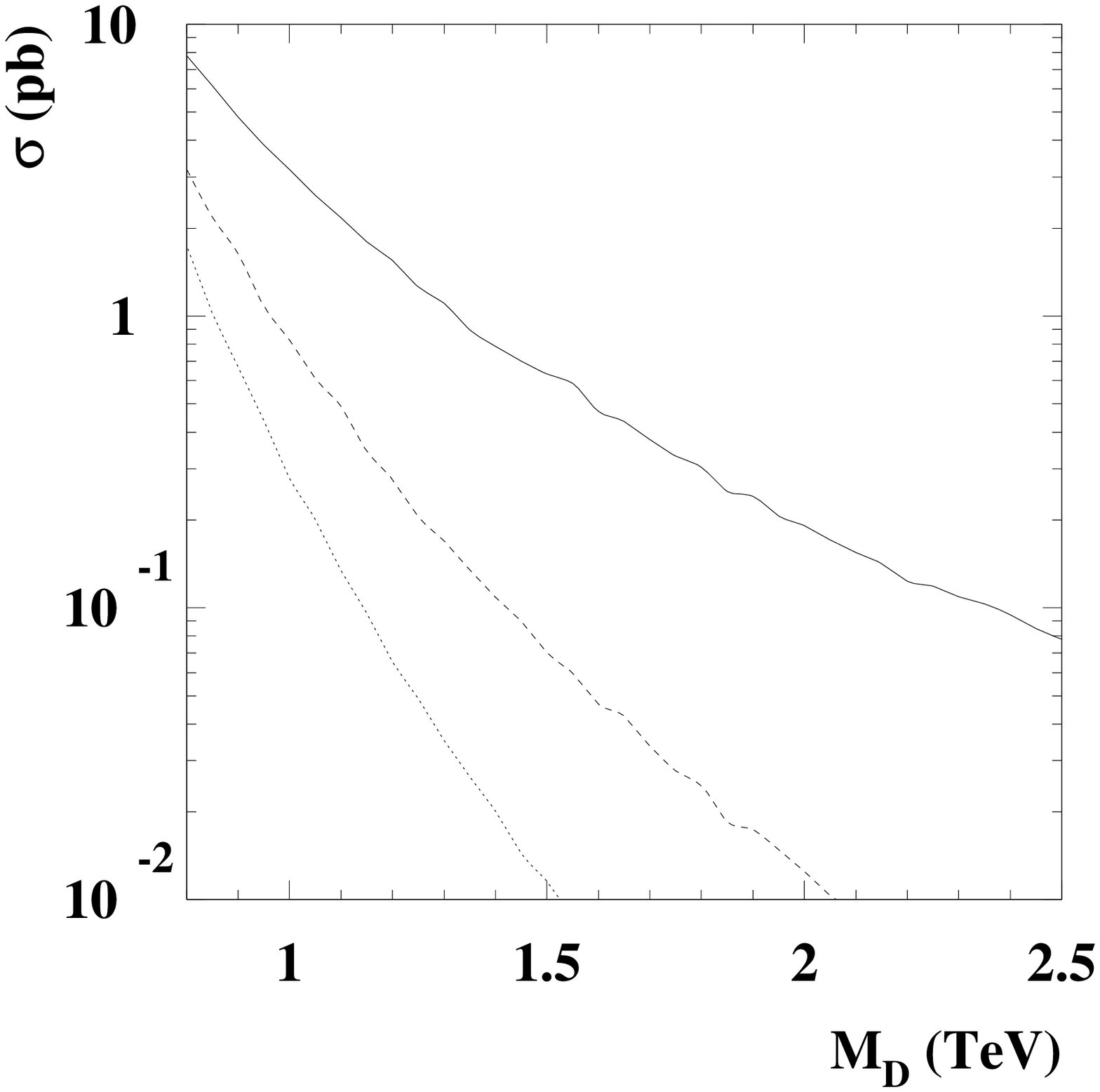}
   \includegraphics[height=3.in]{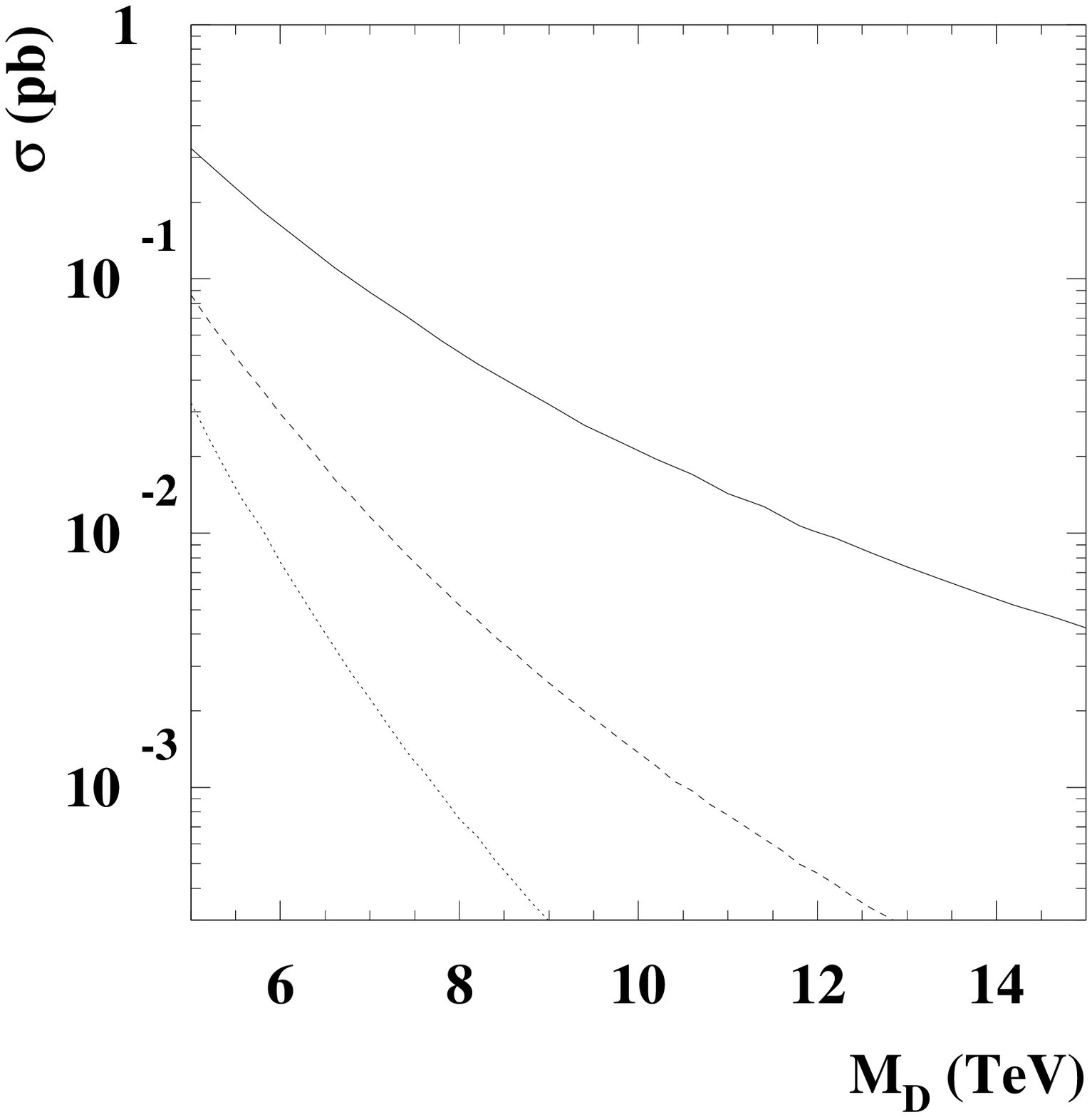}
}
\caption {The jet+ missing energy cross section from graviton and SM quark or
gluon production at Tevatron Run II (left panel) and LHC (right panel). The
solid, dashed and dotted lines correspond to $N=2, 4$ and 6 extra dimensions.}
\label{sm_prod}
\end{figure}

In Fig. \ref{sm_prod} we present the jet + $\not{E_T}$ cross
section as a function of the fundamental gravity scale $M_D$ for
Tevatron Run II and LHC. Cuts on the jet transverse momentum ($p_T
> 200$ GeV at Tevatron, $p_T > 1$ TeV at LHC) and rapidity ($|y| <
3.0$) have been used. We estimate the SM background (assumed to
come only from physics, jet + $Z$ production, with $Z$ decaying to
neutrinos) at parton level, using MADEVENT \cite{Madgraph},
 and with these cuts we obtain around
0.14 pb for Tevatron and 10 fb for LHC. The results are similar
with the pure ADD cross-sections presented in
\cite{peskin-add,GRW}, but for a numerical comparison one should
note that the definition for $M_D$
 we use is slightly different (in Eq (\ref{grv-rel}),
a factor of $1/8\pi$ appears on the left hand side in \cite{GRW}, while
$1/4\pi$ appears on the left hand side in \cite{peskin-add}).

\begin{figure}[t!] 
\centerline{
   \includegraphics[height=3.in]{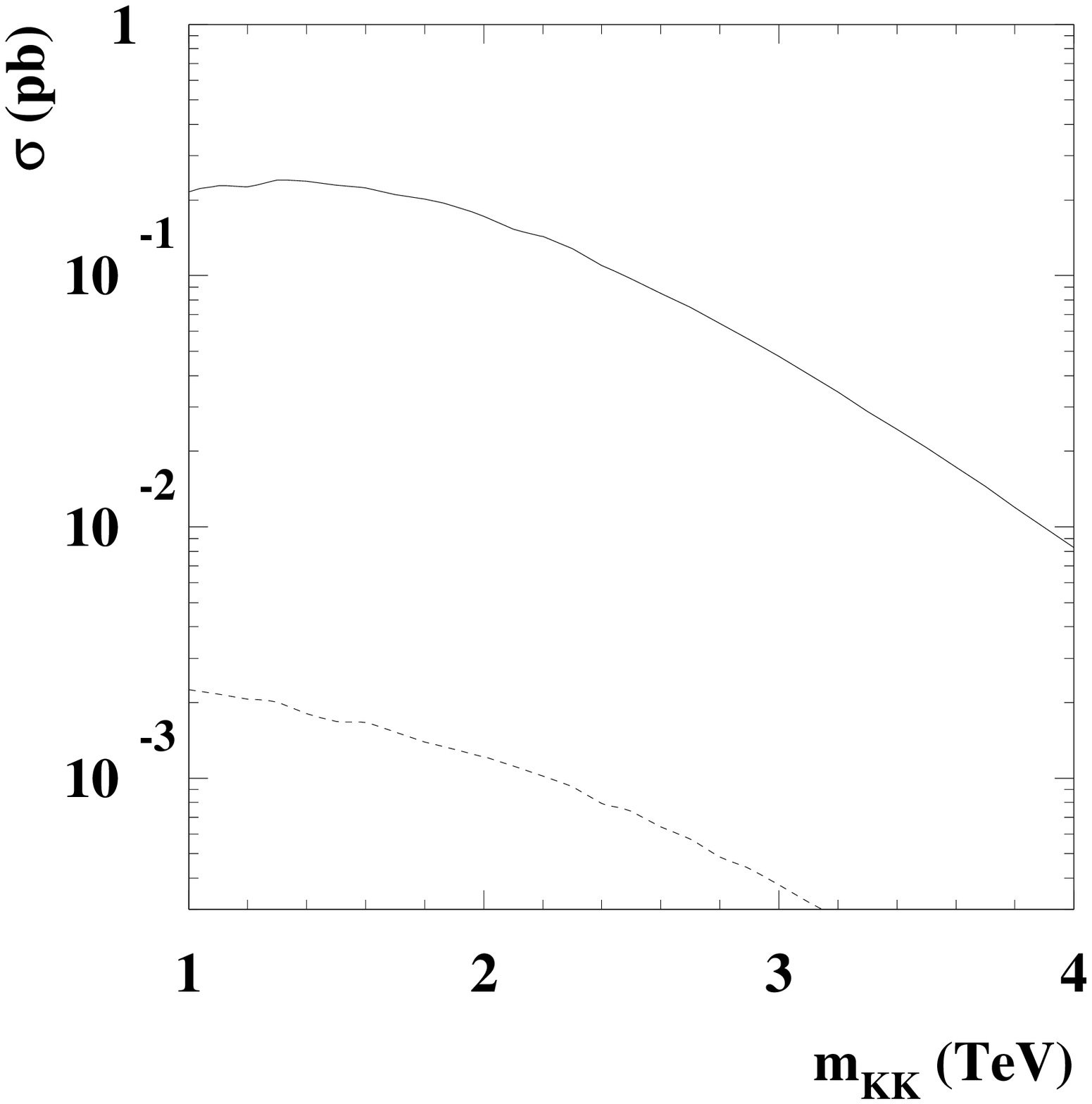}
   \includegraphics[height=3.in]{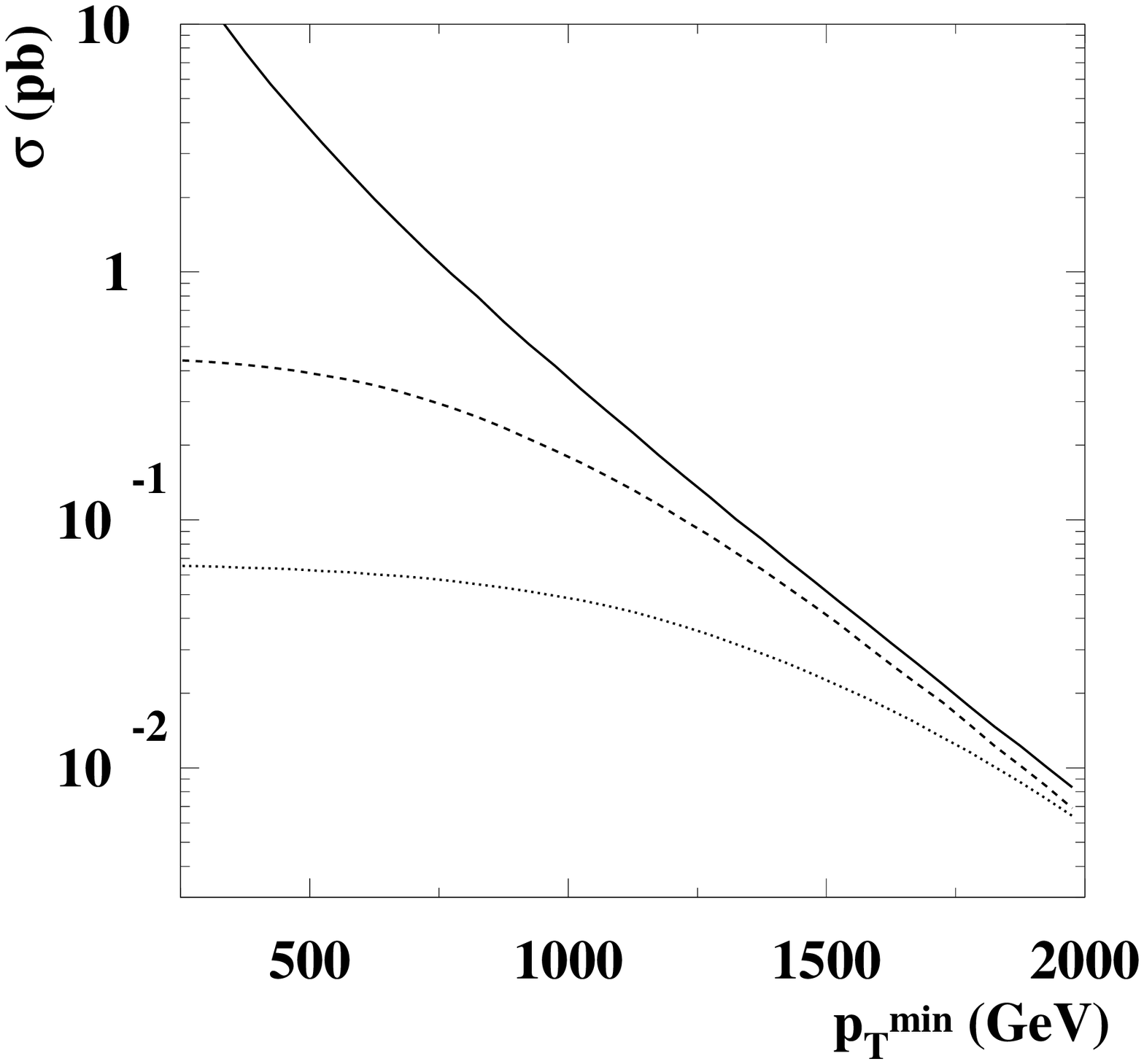}
}
\caption {Left panel: the jet+ missing energy cross section from graviton
and KK quark or gluon production at LHC. The
solid  line corresponds to $N=2$, while the  dashed  line corresponds to $N=4$.
Right panel: the distribution of the cross-section as a function of the cut
imposed on the jet transverse momentum, for production of SM particle
(solid line), and production of KK excitations with $m_{KK}= $2 TeV
(dashed line) and $m_{KK}= $3 TeV
(dotted line).}
\label{kk_prod_lhc}
\end{figure}

In Fig. \ref{kk_prod_lhc} (left panel) we present the cross section
for the production of
one quark or gluon KK excitation and a graviton at the LHC, shown as
a function of the mass  of the KK particle $m_{KK}$. The same cuts
as for the case of SM particle production are used;
we have used a value of $M_D$ = 5 TeV for this plot.  We see that
while the production cross-section may be large enough in some
cases (especially for a small number of extra dimensions) for this
signal to be measurable, it is generally smaller than the signal
due to production of the SM particle together with graviton. We
also note that the cross-section is somewhat flat for small values
of $m_{KK}$. This behavior is due to the large $p_T$ cut imposed
on the momentum of the observable jet. Since in this case the jet
comes from the decay of a massive particle (the KK excitation),
its transverse momentum will tend to increase with the mass of the
particle. This partially compensates for the decrease in
cross-section due to the production of heavier particles.

In Fig. \ref{kk_prod_lhc} (right panel) we present the cross
section as a function of the cut $p_T^{min}$ on the transverse
momentum for the observable jet. The solid line corresponds to the
case of SM particle production, while the dashed and dotted line
correspond to the case of gluon/quark KK excitations in the final
state. (with $m_{KK} = 2$ TeV for the dashed line and  $m_{KK} =
3$ TeV for the dotted line). The plot is made for $N=2$ extra
dimensions, with $M_D$ = 5 TeV. As noted above, for small values
of $p_T^{min}$, the cross-section for the production of the SM
particles dominates (this is a consequence of the form-factor
${\cal F}^c_{01|n}$ multiplying the amplitude rather than the
appearance of a massive KK particle in the final state). However,
the transverse momentum of the jets due to production of a SM
particle falls faster than the $p_T$ of the jets coming from the
decay of the heavy KK excitation, and as it can be seen in figure,
at very large $p_T$ the two signals are of comparable magnitude.
This suggests that if one wants to look for the production of KK
excitations of matter in this channel, one should look primarily
at very high $p_T$ events.

However, a better possibility of identifying KK particle
production at hadron colliders will appear if these excitations
decay first to the LKP ($\gamma^*$), which in turn decays
gravitationally. The signal in this case will be a high $p_T$
photon in the final state plus missing energy (there will also be
some soft jets and leptons, but we will not consider those in our
analysis). The SM background in this case is much smaller that for
the case of a jet + $\not{E_T}$; the signal due to the production
of a SM photon with a graviton will also be smaller, since the
production process for such a signal will be an electroweak
process rather than a strong one. For example, at the LHC, for a
$p_T$ cut of 500 GeV, the SM background is $\sim $ 1 fb. With a
100 fb$^{-1}$ of integrated luminosity, a $5 \sigma$ discovery
would then require 50 signal events, or a cross-section of 0.5 fb.
From direct production of a SM photon with a graviton, the values
of $M_D$ for which the cross-section will reach 0.5 fb will be 5.4
TeV for $N=6$, 6 TeV for $N=4$ and 8.3 TeV for $N=2$.

\begin{figure}[b!] 
\centerline{
   \includegraphics[height=3.in]{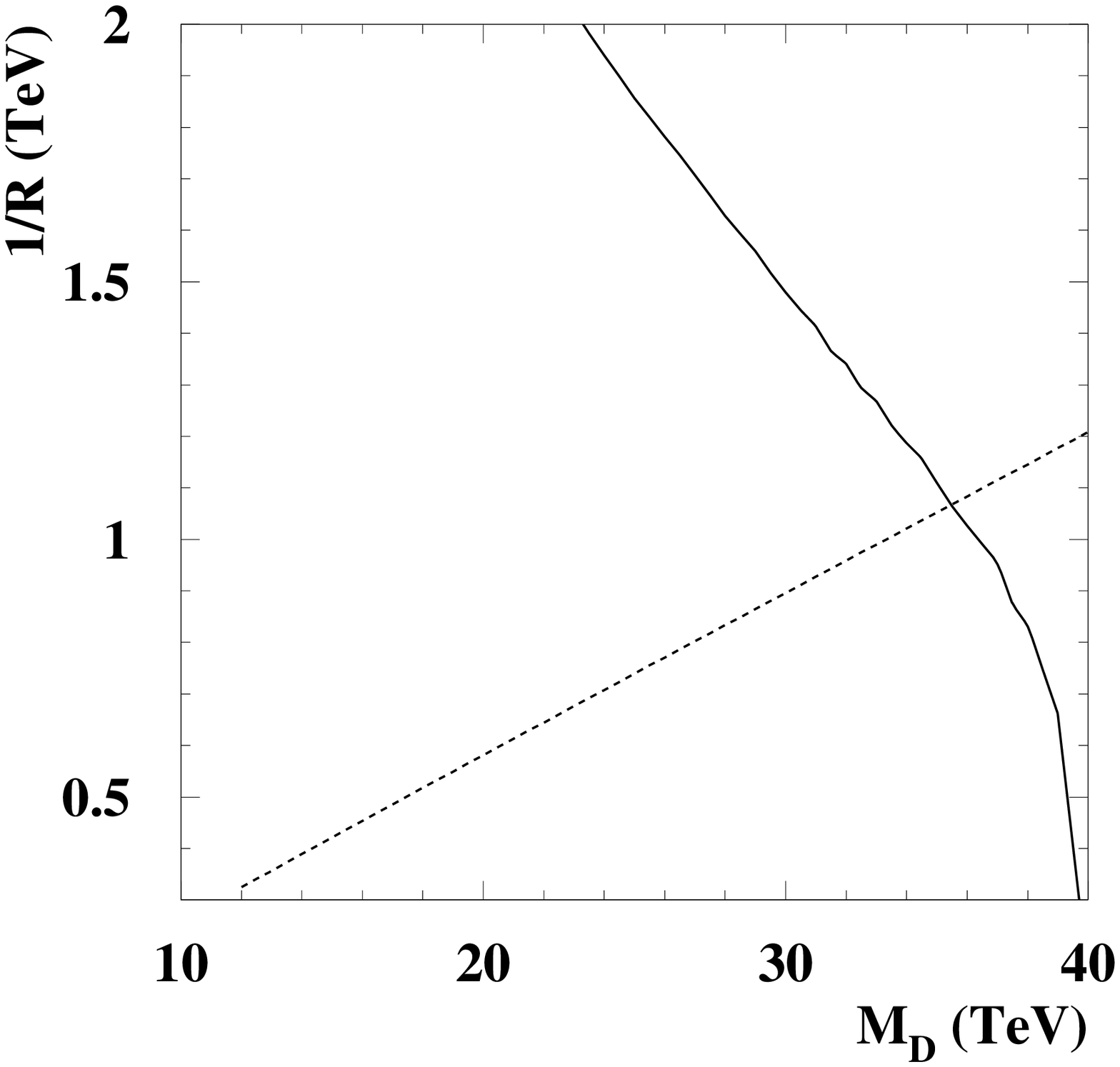}
   \includegraphics[height=3.in]{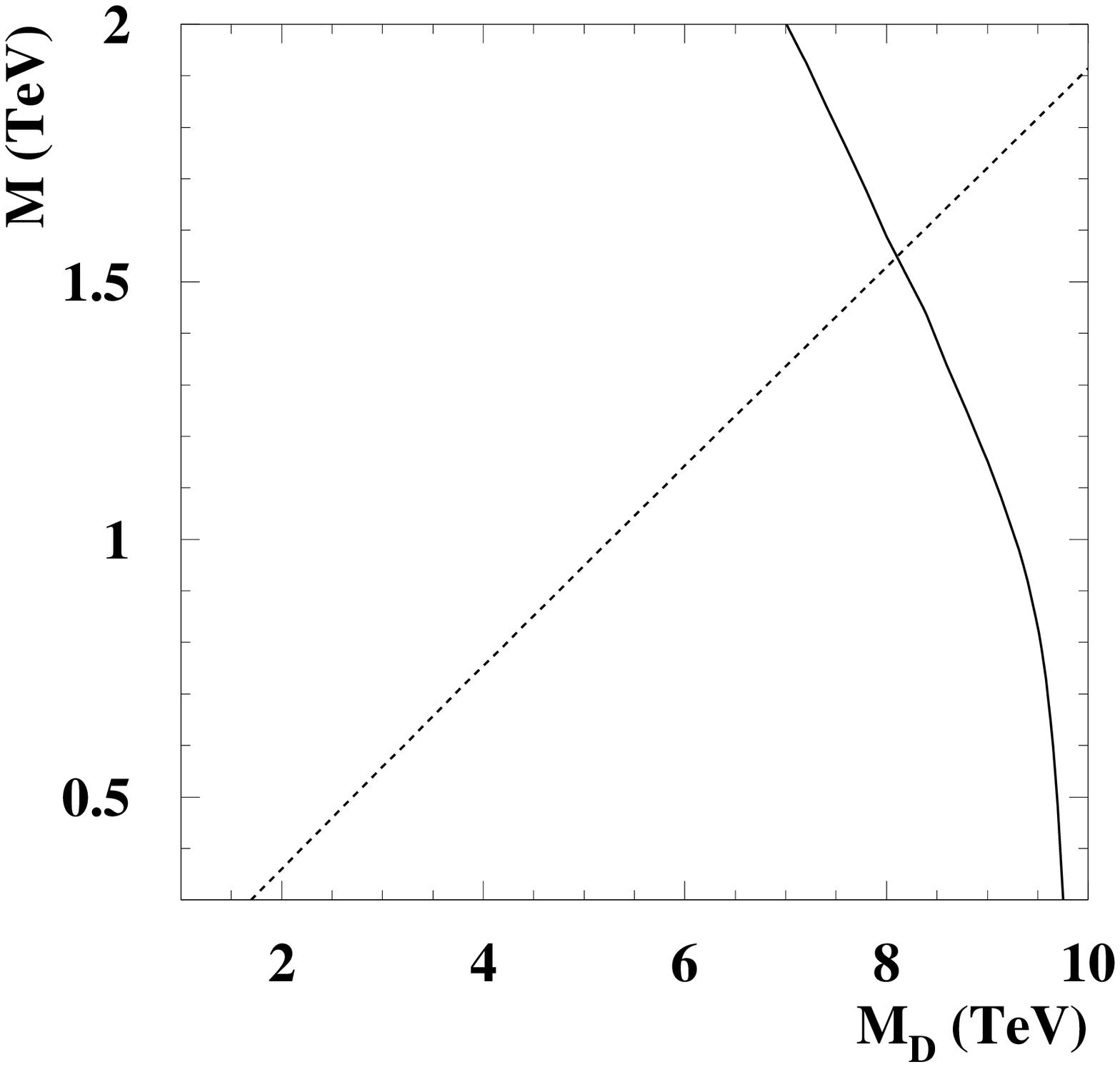}
}
\caption {Solid lines: the $5 \sigma$ discovery reach at the LHC in the
photon + $\not{E_T}$ channel
for $N=2$ (left panel) and $N=4$ (right panel). For values of $M_D, 1/R$
below the dashed lines, the KK quarks and gluons decay first to the LKP.}
\label{reach_n24}
\end{figure}

\begin{figure}[b!] 
\centerline{
   \includegraphics[height=3.in]{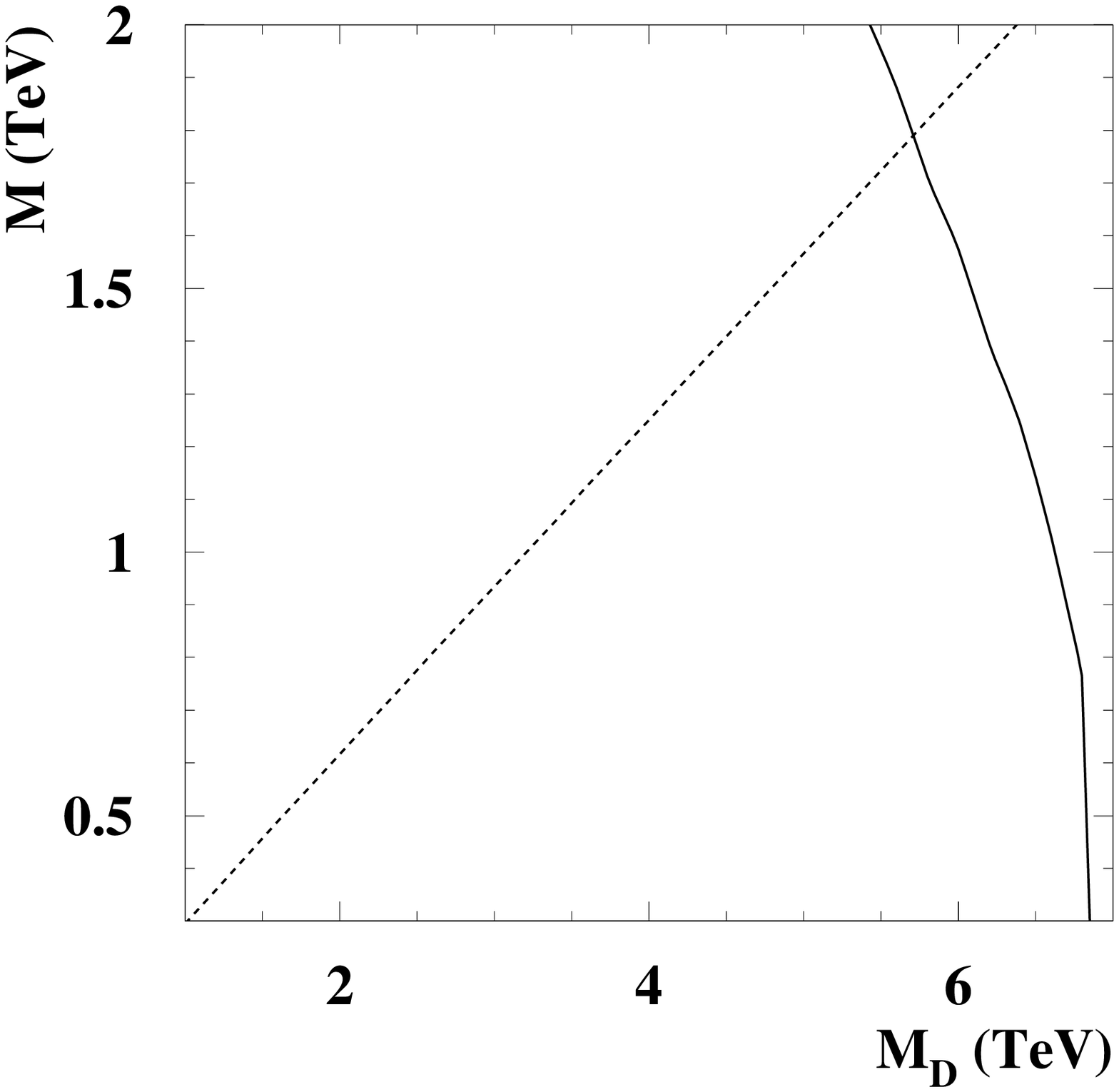}
   \includegraphics[height=3.in]{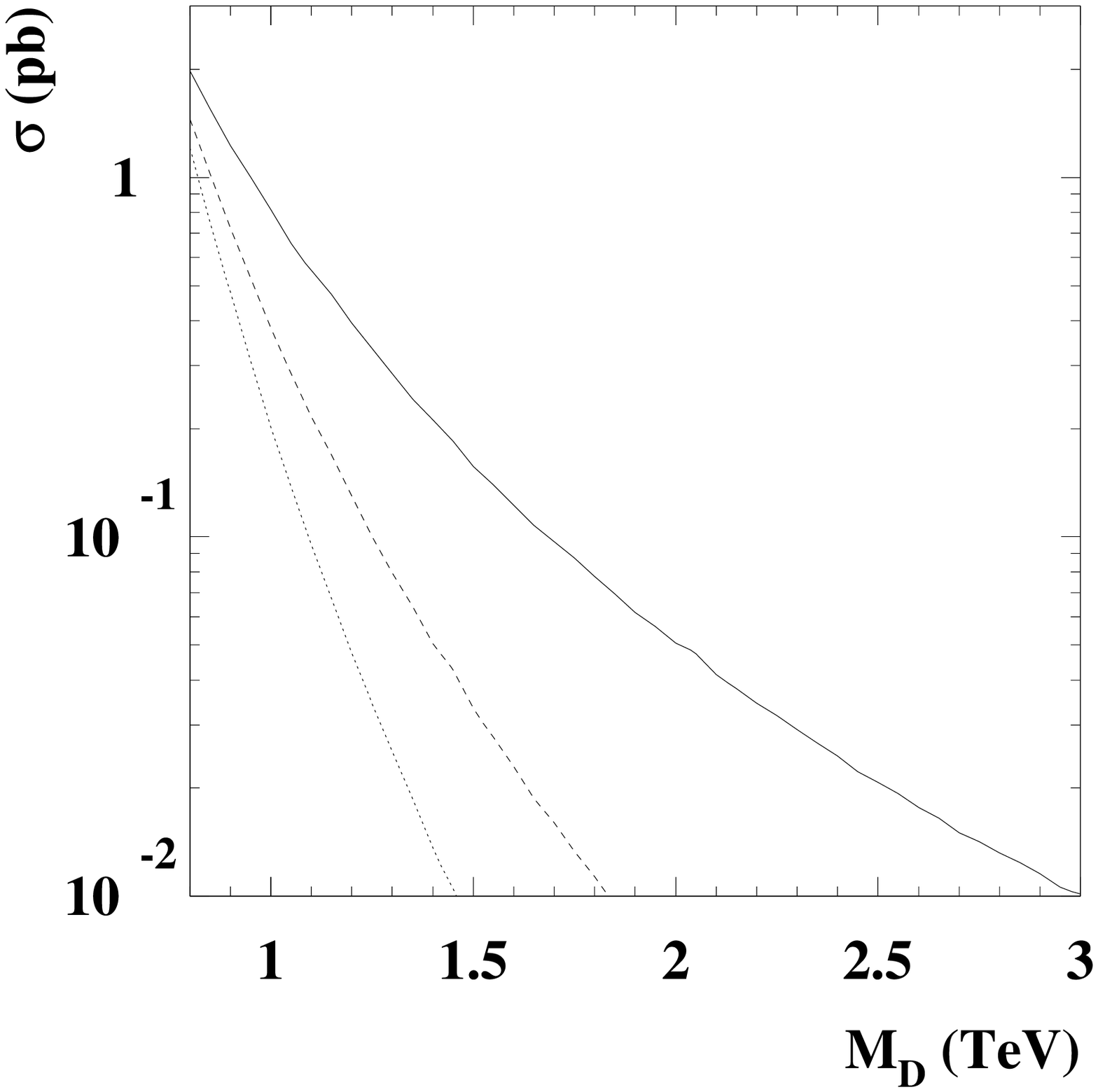}
} \caption {Left panel: the $5 \sigma$ discovery reach at the LHC
in the photon + $\not{E_T}$ channel for $N=6$. Right panel: the SM
photon + $\not{E_T}$ cross-section at the Tevatron Run II, with
$p_T$ $>$ 100 GeV (solid, dashed and dotted lines correspond to
$n=2, 4$ and 6 extra dimensions respectively). } \label{reach_n6}
\end{figure}

In the case of production of a gluon/quark KK excitation which
subsequently decays to a photon, the values of $M_D$ corresponding
to a 0.5 fb cross-section can be as high as $\sim$ 7 TeV for
$N=6$, 10 TeV for $N=4$ and 40 TeV for $N=2$, depending on the
value of $1/R$. We show in Figs. \ref{reach_n24},
\ref{reach_n6}(left panel) with the solid lines the  discovery
reach in the $(M_D,1/R)$ plane (that is, for points below and to
the left of the solid lines, the cross-section will be bigger than
0.5 fb). The dashed lines correspond to values of parameters for
which the gravitational decay widths start becoming important;
that is, for points below the dashed lines the quark and gluon KK
excitations will decay predominantly to $\gamma^*$, while for
points above they will decay directly to SM quarks and gluons
through graviton radiation. Therefore, in the region below both
the solid and dashed lines, the signal will be photon +
$\not{E_T}$, and it will be large enough to ensure discovery. We
see that in this channel we can probe values of $M_D$ similar to
those achievable in the jet + $\not{E_T}$ channel for $N=4$ and
$N=6$ (from Fig. \ref{sm_prod}) and almost twice as large for
$N=2$ (the $5 \sigma$ discovery reach from jet + missing energy
being $\sim 20$ TeV in this case).

The case for the Tevatron is somewhat different.  As it can be
seen from Figs. \ref{reach_n24}, \ref{reach_n6}, the requirement
that $g^*$ and $q^*$ decay first to the LKP implies rather large
values of $M_D$ ($>$ 10 TeV for $N=2$). This means that the
production cross-section is highly suppressed. Typically, for
$N=4,6$ there is just a small region of values for $M_D$ where
$g^*$ and $q^*$ both decay to $\gamma^*$ and are produced in
enough numbers to be observable. However, in that region the
signal coming from production of SM photons plus KK gravitons is
predominant. We show in Fig. \ref{reach_n6}(right panel) the
cross-section for SM photon production plus  $\not{E_T}$ (from KK
gravitons), with a cut on photon $p_T$ of 100 GeV. The Standard
Model background (from $Z \gamma$ production) is $\sim$ 80 fb. It
is interesting to note that the the $M_D$ discovery reach at the
Tevatron is roughly similar in the jet + $\not{E_T}$ channel
versus photon + $\not{E_T}$ channel. The reason is that while the
production cross-section is suppressed by a factor
$\alpha_{em}/\alpha_s \sim 1/10$, the background is similarly
suppressed, and one can use softer cuts (using a $p_T$ cut of 100
GeV as opposed to 200 GeV will increase the cross-section by a
factor of 10). Moreover,  the predominant initial state
responsible for the production cross-section is $q \bar{q}$, which
favors processes with photons in the final state (whereas for the
LHC is $qg$ and $gg$, resulting in additional suppression of final
state photons).

\section{Conclusions}

The scenario of a universe with more than four dimensions leads to
interesting and testable experimental consequences. In this paper
we studied a particular class of models, where the matter fields
 propagate on a 4+1 dimensional fat brane embedded in a 4+$N$
higher dimensional space.
Gravity propagates in the whole bulk, with dimensions of order inverse eV,
which helps solve the hierarchy problem,
while the thickness of the brane on which matter propagates
is of order inverse TeV.

The standard phenomenology associated with UED models is affected
by the  inclusion of  the gravitational interactions. These break
KK number conservation, and therefore allow the decays of the  KK
excitations of matter, as well as single production of KK
excitations. In this paper we consider signals with final states
containing a KK graviton. These can be produced together with a SM
quark or gluon (as in the ADD case, when matter is restricted to
the 4D brane), as well as together
 with the KK excitations of such particles.
The phenomenological signal in the later case
depends on the decay of the $q^*/g^*$ excitation, and can be either
be a jet with missing energy (if gravitational decays dominates), or
photon plus missing energy (if the $q^*/g^*$ decays first to the
$\gamma^*$).


We find that, for final states  consisting of a monojet plus
missing energy, the production cross-section for SM particle plus
a graviton (similar to that obtained  in pure ADD models) is
typically larger than the one for KK particle plus graviton.
However, for the case of $N=2$ extra dimensions, in regions with
large transverse momentum the signal associated with the
production of a KK excitation is comparable in magnitude to the
signal resulting from the production of a SM particle. In our
model, using this channel one can then probe the fundamental scale
of gravity $M_D$ starting from 1.5 TeV up to 3 TeV at the
Tevatron, and starting from 8 TeV up to 25 TeV at the LHC (the
lower values correspond to $N=6$, while the higher values
correspond to $N=2$ extra dimensions).

Things are somewhat different for the high $p_T$ photon plus
missing energy channel. The cross-section due to SM photon
production is reduced due to the appearance of the electroweak
coupling constant, as well as due to the requirement that the
initial state is $q \bar{q}$ (which affects mostly the LHC case).
One then finds at the LHC, for sufficiently small values for the
mass of matter KK excitations $M$ (such that these decay first to
the LKP $\gamma^*$), the signal associated with production of KK
matter dominates. One can probe values of $M_D$ starting from 7
TeV (for $N = 6$) up to 40 TeV (for $N=2$). The larger reach in
the $N=2$ case compared to the jet + $\not{E_T}$ final state is
due to lower background in the photon channel, as well as to the
enhancement of the production cross-section for light gravitons in
the final state. On the other hand, for values of $M_D$ accessible
at Tevatron, the quark and gluon excitations will decay mostly to
gravitons plus jets, and the dominant source of final state
photons will be due to the production of the SM particles. Again,
due to smaller backgrounds, the discovery reach for $M_D$  in this
channel is comparable to the one obtained in the jet + $\not{E_T}$
channel.


\end{document}